\documentclass[12pt,a4paper]{article} 

\usepackage{amsmath}
\usepackage{amssymb}
\usepackage{mathtools}
\usepackage{gensymb}
\usepackage{bbm, dsfont}
\usepackage[utf8]{inputenc}

\usepackage{dsfont}

\usepackage{caption}
\usepackage{subfig}
\usepackage{booktabs}

\usepackage[margin=3cm]{geometry}

\usepackage{longtable}
\usepackage{booktabs}
\usepackage{tabu}

\usepackage{authblk}

\usepackage[dvipsnames]{xcolor}

\usepackage{siunitx} 
\usepackage{tabularx} 
\newcolumntype{A}{>{\centering\arraybackslash}X}
\newcolumntype{L}[1]{>{\raggedright\arraybackslash}p{#1}}
\newcolumntype{C}[1]{>{\centering\arraybackslash}p{#1}}
\newcolumntype{R}[1]{>{\raggedleft\arraybackslash}p{#1}}

\newcommand{\ackname}{Acknowledgements}

\begin{document}

\title{Complete list of Bell inequalities with four binary settings}

\author[1]{E. Zambrini Cruzeiro\thanks{emmanuel.zambrinicruzeiro@unige.ch}}
\author[1]{N. Gisin}
\affil[1]{Group of Applied Physics, University of Geneva, 1211 Geneva, Switzerland}
\date{}                     
\setcounter{Maxaffil}{0}
\renewcommand\Affilfont{\itshape\small}

\maketitle

\begin{abstract}
We give the complete list of 175 facets of the local polytope for the case where Alice and Bob each choose their measurements from a set of four binary outcome measurements. For each inequality we compute the maximum quantum violation for qubits, the resistance to noise, and the minimal detection efficiency required for closing the detection loophole with maximally entangled qubit states, in the case where both detectors have the same efficiency (symmetric case).
\end{abstract}

\section{Introduction}

Bell inequalities are central to the study of non-locality, but finding the complete list of Bell inequalities for a given Bell scenario can be a very difficult task \cite{Pitowski1989}. A Bell scenario is specified by a number of measurement settings and a number of measurement outcomes for each party. In the case of two parties with two measurement choices each (the simplest case), there is only one Bell inequality, the Clauser-Horne-Shimony-Holt (CHSH) inequality \cite{Clauser1969}. The local polytope has two facets, CHSH and positivity. If one allows both parties to choose between three binary outcome measurements, there is only one new relevant inequality besides CHSH. For four settings on each side, the number of facet inequalities grows to 175, where 169 of these inequalities genuinely use the four settings. The complete list of inequalities has not been known until recently, thanks to the research behind \cite{Brunner2008,Pal2009,Bancal2010}, but was never presented. Therefore one could find an almost complete list of the inequalities distributed in the literature. As a service to the community, we present the complete list in a single document. In addition, we study the basic quantum properties of these inequalities by computing the local and two-qubit quantum bounds, the state which attains the quantum bound, the resistance to noise for both the state that violates maximally the inequality and the maximally entangled state, and finally the minimum detector efficiency required to close the detection loophole assuming Alice and Bob's detectors have the same efficiency and the maximally entangled two-qubit state. 

In Section \ref{sec:one}, we review all Bell inequalities for scenarios with fewer binary-outcome measurements. In Section \ref{sec:two} we describe the computation of the local and quantum bounds, the resistance to noise and minimal detection efficiency to close the detection loophole, and present the main results.

\section{Review of all Bell inequalities with less settings}
\label{sec:one}

In any bipartite Bell scenario, the statistics are fully specified by the joint probability distribution $p(ab|xy)$, where $a,b$ and $x,y$ are the outputs and inputs of Alice and Bob, respectively. For binary outcome scenarios XY22, where X is the number of settings of Alice and Y the number of settings of Bob, there are 4XY probability elements that specify $p(ab|xy)$. Some of these elements are not necessary to fully specify the statistics though, as they are not independent of each other due to the normalization and non-signalling conditions. Taking these into account, one finds that there are only 4XY-XY-X(Y-1)-Y(X-1)=XY+X+Y independent probability elements. Therefore one can fully specify the statistics of a binary outcome Bell test using XY+X+Y elements, this is the idea behind the Collins-Gisin (CG) notation \cite{Collins2004}. As we deal only with binary outcome measurement, in the rest of the paper we shall denote a XY22 scenario simply by XY.

Using the CG notation, the probability distributions of 22 are specified by a table with the following elements

\begin{table}[!htbp]
  \begin{center}
    \label{tab:notation}
    \begin{tabular}{l|c c} 
       & $p^B(0|0)$ & $p^B(0|1)$ \\
      \hline
      $p^A(0|0)$ & $p(00|00)$ & $p(00|01)$\\
      $p^A(0|1)$ & $p(00|10)$ & $p(00|11)$\\
    \end{tabular}$\leq$ b
  \end{center}
\end{table}
\noindent where $p^A(a|x)$ and $p^B(b|y)$ are the marginals of Alice and Bob respectively.

Such a table defines a joint probability distribution $p(ab|xy)$. One can also describe an inequality using such a table, in this case each element of the table instead represents the coefficient that multiplies the probability element indicated in the probability table. We introduce coefficients for the joint probability distribution $d_{xy}$, the marginals of Alice $c_x$ and those of Bob $e_y$. In general, a Bell inequality for a XY scenario can be written as

\begin{equation}
I=\sum_{xy}d_{xy}p(00|xy)+\sum_xc_xp^A(0|x)+\sum_ye_yp^B(0|y)\leq b
\end{equation}

%

In the 22 scenario the only relevant inequality is the CHSH inequality
\begin{table}[!htbp]
  \begin{center}
    \label{tab:chsh}
    CHSH=\begin{tabular}{l|c c} 
       & -1 & 0 \\
      \hline
      -1 & 1 & 1\\
      0 & 1 & -1\\
    \end{tabular}$\leq$ 0
  \end{center}
\end{table}

As described in the introduction, even for the case of binary outcomes few is known. All tight Bell inequalities are known for the following cases: 22, X2 (X $\leq 3$), 33 and 43 and 34. For the cases 22 and X2 (X $\leq 3$), there is only one Bell inequality \cite{Collins2004,Fine1982}: the CHSH inequality.

The 33 scenario has one new inequality \cite{Collins2004},

\begin{table}[!htbp]
  \begin{center}
    \label{tab:I3322}
    $I_{3322}=$\begin{tabular}{l|c c c} 
       & -1 & 0 & 0 \\
      \hline
      -2 & 1 & 1 & 1\\
      -1 & 1 & 1 & -1\\
      0  & 1 & -1 & 0
    \end{tabular}$\leq$ 0
  \end{center}
\end{table}

If Alice's third setting and Bob's first setting are not used, $I_{3322}$ reduces to CHSH. Therefore in terms of minimal detection efficiency (symmetric) to close the detection loophole, $I_{3322}$ and CHSH perform the same.

In the 34 scenario, there are three new inequivalent inequalities

\begin{table}[!htbp]
  \centering
  \subfloat[][]{$I^{(1)}_{3422}=$\begin{tabular}{l|c c c c} 
       & 1 & 0 & 0 & 1 \\
      \hline
      1 & -1 & -1 & 1 & -1\\
      1 & -1 & 1 & -1 & -1\\
      -2 & 1 & 1 & 1 &-1\\
    \end{tabular}$\leq$ 2}%
  \qquad
  \subfloat[][]{$I^{(2)}_{3422}=$\begin{tabular}{l|c c c c} 
       & -1 & 0 & -1 & 1 \\
      \hline
      0 & -1 & 0 & 1 & -1\\
      1 & 1 & -1 & 0 & -1\\
      -1 & 1 & 1 & 1 & 0\\
    \end{tabular}$\leq$ 1}
\qquad
  \subfloat[][]{$I^{(3)}_{3422}=$\begin{tabular}{l|c c c c} 
       & 0 & 0 & -1 & 2 \\
      \hline
      1 & -2 & 0 & 1 & -1\\
      0 & 1 & -1 & 1 & -1\\
      -1 & 1 & 1 & 1 & -1\\
    \end{tabular}$\leq$ 2}
  \label{tbl:I3422}%
\end{table}

The optimal settings for $I_{3422}^{(1)}$ assuming the maximally entangled state are the same optimal settings of the Elegant Bell inequality \cite{Gisin2009,Gisin2017}. Alice's optimal settings are three orthogonal measurements X,Y and Z, while Bob's settings form the vertices of a regular tetrahedron on the Bloch sphere. With $I_{3422}^{(1)}$ one can have a larger quantum violation using partially entangled two-qubit states. In this case, the optimal settings of Alice are still X,Y,Z, but Bob's settings become an irregular tetrahedron.

All the inequalities we present in this section can be lifted to the 44 scenario \cite{Pironio2005}, therefore we already know five inequalities of 44. These five liftings correspond to inequalities 1 to 5 in Table \ref{tab:one}. The first 31 inequalities of Table \ref{tab:one} are the 31 inequalities published in \cite{Brunner2008}, given in the same order. The list of coefficients in the CG notation for each inequality is provided separately as a file, and the inequalities are listed in it in the same order. 


\section{Bell inequalities with four settings for each party}
\label{sec:two}

We give the complete list of 175 facets of the local polytope in the case of four binary outcome measurements for both Alice and Bob. We present all the inequalities in a file, except the trivial one (positivity: $p(ab|xy)\geq 0$ for all $a,b,x,y$), in the case of four outcomes for Alice and Bob. The main results are given in Table \ref{tab:one}. Along with each inequality, we provide the local bound $L$, quantum bound $Q$, the resistance to noise $\lambda$, and the minimal detection efficiency $\eta$ required for closing the detection loophole in the symmetric case. Note that six of the 175 inequalities correspond to liftings: positivity, CHSH, $I_{3322}$, $I_{3422}^{(1)}$, $I_{3422}^{(2)}$ and $I_{3422}^{(3)}$. Therefore there are 169 inequalities which use the four settings on both sides.

\subsection{Quantum violation}

The local bound $L$ is computed by finding the optimal strategy using only shared randomness and local operations. Joint probability distributions $p(ab|xy)$ which are local in this sense can be decomposed in the following way
\begin{equation}
p(ab|xy)=\int q(\lambda)p^A(a|x)p^B(b|y)d\lambda
\end{equation}
where $\lambda$ is the shared random variable and $q(\lambda)$ is its probability distribution.

The quantum bound $Q$ gives the maximal quantum violation for a two-qubit pure entangled state of the form
\begin{equation}
|\psi (\theta)\rangle=\cos\theta |00\rangle +\sin\theta |11\rangle
\end{equation}

Note that we compute the maximal quantum bound for qubits. Nevertheless, it is known that using higher-dimensional states one can in some cases achieve a better quantum bound \cite{Pal2010}. 

We optimize over projective non-degenerate von Neumann measurements. Each measurement setting of Alice (Bob) is described by a vector $\vec{a}_x$ ($\vec{b}_y$) on the Bloch sphere. One has 
\begin{equation}
p(00|xy)=\mathrm{Tr}\left(A_x\otimes B_y |\psi (\theta)\rangle\langle\psi (\theta)|\right)
\end{equation}
where $A_x\coloneqq \frac{1+\vec{a}_x\cdot\vec{\sigma}}{2}$, and similarly for $B_y$.

All the inequalities for scenarios 22, 32, 23 and 33 are maximally violated by maximally entangled qubit states. This is not the case for scenarios 34 and 44, where the maximal violation can be given by a partially entangled state. In the 44 scenario, we find inequalities which are maximally violated by partially entangled states for non-degenerate measurements, which can be in some cases very far from the maximally entangled state. An example of such an inequality is facet number 15, which is maximally violated by the state
\begin{equation}
\psi(\theta_{\mathrm{max}})=0.4018|00\rangle + 0.9157|11\rangle
\end{equation}
The violation is small compared to other inequalities of this table, and we see that the resistance to noise is bad. However, as described in \cite{Brunner2008}, using degenerate measurements this inequality ($I_{4422}^4$ in \cite{Brunner2008}) is maximally violated by the maximally entangled state. Indeed using degenerate measurements this inequality becomes equivalent to CHSH.

\subsection{Resistance to noise}

Let $|\psi\rangle$ be the state that maximally violates a specific bell inequality. Then the resistance to noise $1-\lambda$ is defined as the amount of white noise that can be mixed with $|\psi\rangle$ in order for the bell inequality not to be violated anymore.

\begin{equation}
\rho=\lambda|\psi\rangle\langle\psi | +(1-\lambda)\frac{\mathds{1}}{4}
\end{equation}

The best bipartite inequality XY for X,Y $\leq 4$ in terms of resistance to noise using qubits is by far CHSH.

\subsection{Minimum detection efficiency for closing the detection loophole}

There are different possible strategies to close the detection loophole. All of them involve preventing the attacker (Eve) from exploiting the non-detections. How should the experimenter handle a non-detection event? One possible strategy, which was implemented in \cite{Brunner2008}, is that Alice and Bob output $a=0$ or $b=0$ respectively, in order to deal with a non-detection. However, they could also output 1, giving four possibilities in total. We have also optimized the detection efficiencies over these no-click strategies.

A Bell inequality with detector efficiency $\eta$ for both Alice and Bob can be written \cite{Brunner2007}:
\begin{equation}\label{eq:polynomial}
I_{\eta,\eta}=\eta^2 Q + \eta (1-\eta)(M_A+M_B)+(1-\eta)^2X\leq L
\end{equation}
where $M_A$ is the value that the Bell inequality yields when only Alice's detector fires, $M_B$ is the bound when only Bob's detector fires and $X$ accounts for when both detectors do not fire.

We find that the smallest value for the symmetric detection efficiency using a maximally entangled two-qubit state is $82.14\%$, which is already the best result of \cite{Brunner2008}. This inequality, labeled $A_5$ in \cite{Brunner2008}, is number 8 on our table. 

\subsection{Correlation inequalities}

In this section, we present the two inequalities of this list which can be cast into correlator-only form. A correlator $E(x,y)$ is defined as
\begin{equation}
E(x,y)=p(a=b|xy)-p(a\neq b|xy)
\end{equation}
CHSH for example can be put into full-correlation form
\begin{equation}
E(0,0)+E(0,1)+E(1,0)-E(1,1) \overset{L}{\leq} 2 \overset{Q}{\leq} 2\sqrt{2}
\end{equation}
where $L$ and $Q$ relate to the local and quantum bounds, respectively.

We find that only three inequalities can be put into full-correlation form: inequalities 1 (CHSH), 10 and 11. Facet number 10 is the inequality named $AS_1$ in \cite{Brunner2008}, while facet number 11 is $AS_2$. These inequalities are the only ones which can be put in full-correlation form for the 44 scenario, which confirms the result of D. Avis et al. \cite{Avis2006}.

\section{Conclusion}

We have studied the complete list of four binary-outcome settings Bell inequalities. We give the full list and a table with the local and quantum bounds of all inequalities, the two-qubit state that maximally violates each inequality, the resistance to noise and the minimal detection efficiency for maximally entangled qubit states to close the detection loophole in a Bell experiment where both detectors have the same efficiency. We find several inequalities which are maximally violated by partially entangled states, which is interesting for the study of nonlocality. It is also confirmed that the minimum detection efficiency is $82.14\%$, and is found for the inequality $A_5$ published in \cite{Brunner2008}.

\fontsize{11}{16}\selectfont

\section*{Added note}
\textit{While this article was being written, the complete list of inequalities was presented in \cite{Oudot2018}, although the inequalities were not studied. In order not to confuse readers, we have added the same name convention for the inequalities that is used in \cite{Oudot2018}.}

\section*{\ackname}
\textit{The authors thank J.D. Bancal for useful discussions. Financial support by the Swiss NCCR-QSIT is gratefully acknowledged.}

\fontsize{12}{16}\selectfont

\bibliography{blind}

\newpage
\newgeometry{margin=0.3cm}
\small
\begin{longtabu}{|c|c|c|c|c|c|c|c|}
\caption{Quantum properties of all Bell inequalities with four binary-outcome settings for both parties. For each inequality we indicate the name under which they can be found in the literature, we give the quantum state that achieves the largest violation $|\psi(\theta_{\mathrm{max}})\rangle=\cos\theta_{\mathrm{max}}|00\rangle +\sin\theta_{\mathrm{max}}|11\rangle$, and we give its resistance to noise $\lambda$. For the maximally entangled state, we provide the resistance to noise $\lambda_{\mathrm{ME}}$, as well as the detection efficiency $\eta$ required to close the symmetric detection loophole. All quantities are computed for two-qubit systems and non-degenerate measurements.}
\label{tab:one}
\\
\hline
\rowfont{\bfseries}
\# & Name & L & Q & $\theta_{\mathrm{max}}/\pi$ & $\lambda$ & $\lambda_{\mathrm{ME}}$ & $\eta_{\mathrm{sym}}$\\
\hline
1 & CHSH & 1 &  1.2071 &  0.25 &  \textbf{0.7071} &  \textbf{0.7071} &  0.8284 \\ 
2 & $I_{3322}$ & 1 &  1.25 &  0.25 &  0.8 &  0.8 &  0.8284 \\ 
3 & $I_{4322}^1$ & 1 &  1.2361 &  0.2332 &  0.864 &  0.866 &  0.8761 \\
4 & $I_{4322}^2$ & 1 &  1.2596 &  0.2251 &  0.828 &  0.8333 &  0.8685 \\ 
5 & $I_{4322}^3$ & 1 &  1.4365 &  0.25 &  0.7746 &  0.7746 &  0.8514 \\ 
6 & $I_{4422}^1$ & 1 &  1.197 &  0.2356 &  0.8988 &  0.9 &  0.8571 \\ 
7 & $I_{4422}^2$ & 2 &  2.6214 &  0.2476 &  0.763 &  0.763 &  0.8443 \\ 
8 & $A_5$ & 1 &  1.4353 &  0.2447 &  0.7751 &  0.7752 &  \textbf{0.8214} \\ 
9 & $A_6$ & 1 &  1.2321 &  0.25 &  0.8829 &  0.8829 &  0.8373 \\ 
10 & $AS_1$ & 3 &  3.5412 &  0.25 &  0.7348 &  0.7348 &  0.8472 \\ 
11 & $AS_2$ & 4 &  4.8785 &  0.25 &  0.74 &  0.74 &  0.8506 \\ 
12 & $AII_1$ & 3 &  3.6056 &  0.2435 &  0.7676 &  0.7679 &  0.8323 \\ 
13 & $AII_2$ & 2 &  2.5 &  0.25 &  0.8 &  0.8 &  0.8508 \\ 
14 & $I_{4422}^3$ & 1 &  1.238 &  0.2257 &  0.863 &  0.866 &  0.8761 \\ 
15 & $I_{4422}^4$ & 1 &  1.056 &  \textbf{0.1315} &  0.9728 &  1 &  1 \\ 
16 & $I_{4422}^5$ & 2 &  2.4365 &  0.25 &  0.7746 &  0.7746 &  0.8514 \\ 
17 & $I_{4422}^6$ & 1 &  1.4495 &  0.25 &  0.8165 &  0.8165 &  0.8697 \\ 
18 & $I_{4422}^7$ & 2 &  2.4548 &  0.2379 &  0.7937 &  0.7949 &  0.8405 \\ 
19 & $I_{4422}^8$ & 3 &  3.4207 &  0.2456 &  0.856 &  0.8561 &  0.8893 \\ 
20 & $I_{4422}^9$ & 2 &  2.4617 &  0.2352 &  0.8441 &  0.8455 &  0.8392 \\ 
21 & $I_{4422}^{10}$ & 2 &  2.6139 &  0.2461 &  0.8175 &  0.8176 &  0.8458 \\ 
22 & $I_{4422}^{11}$ & 3 &  3.6384 &  0.2444 &  0.779 &  0.7792 &  0.8474 \\ 
23 & $I_{4422}^{12}$ & 3 &  3.6188 &  0.2404 &  0.7843 &  0.7849 &  0.8382 \\ 
24 & $I_{4422}^{13}$ & 1 &  1.25 &  0.2069 &  0.8889 &  0.8889 &  0.8944 \\ 
25 & $I_{4422}^{14}$ & 2 &  2.4103 &  0.238 &  0.8298 &  0.831 &  0.8523 \\ 
26 & $I_{4422}^{15}$ & 1 &  1.25 &  0.25 &  0.8889 &  0.8889 &  0.8944 \\ 
27 & $I_{4422}^{16}$ & 1 &  1.2407 &  0.219 &  0.8791 &  0.8829 &  0.9009 \\ 
28 & $I_{4422}^{17}$ & 3 &  3.6714 &  0.2488 &  0.7883 &  0.7883 &  0.8611 \\ 
29 & $I_{4422}^{18}$ & 2 &  2.1812 &  0.2064 &  0.9508 &  0.9623 &  0.9575 \\ 
30 & $I_{4422}^{19}$ & 3 &  3.4307 &  0.25 &  0.8745 &  0.8745 &  0.887 \\ 
31 & $I_{4422}^{20}$ & 2 &  2.3056 &  0.25 &  0.9075 &  0.9231 &  0.899 \\ 
32 & $J_{4422}^{44}$ & 1 &  1.4145 &  0.2279 &  0.8085 &  0.8122 &  0.8677 \\ 
33 & $J_{4422}^{25}$ & 2 &  2.4459 &  0.2393 &  0.797 &  0.7977 &  0.8638 \\ 
34 & $J_{4422}^{60}$ & 1 &  1.5923 &  0.25 &  0.7715 &  0.7715 &  0.8351 \\ 
35 & $J_{4422}^{43}$ & 2 &  2.4414 &  0.2287 &  0.8192 &  0.8221 &  0.8788 \\ 
36 & $J_{4422}^{69}$ & 2 &  2.4693 &  0.2423 &  0.8099 &  0.8104 &  0.8581 \\ 
37 & $J_{4422}^{74}$ & 1 &  1.4332 &  0.2432 &  0.8219 &  0.8223 &  0.8603 \\ 
38 & $J_{4422}^6$ & 2 &  2.4158 &  0.2407 &  0.8279 &  0.8284 &  0.8502 \\ 
39 & $J_{4422}^1$ & 2 &  2.3871 &  0.2393 &  0.8378 &  0.8386 &  0.8583 \\ 
40 & $J_{4422}^{26}$ & 3 &  3.6402 &  0.2436 &  0.7785 &  0.7788 &  0.8409 \\ 
41 & $J_{4422}^{63}$ & 2 &  2.6035 &  0.2433 &  0.7885 &  0.789 &  0.8589 \\ 
42 & $J_{4422}^{75}$ & 3 &  3.627 &  0.2405 &  0.7821 &  0.7827 &  0.8641 \\ 
43 & $J_{4422}^{32}$ & 3 &  3.5902 &  0.2386 &  0.7922 &  0.7931 &  0.8565 \\ 
44 & $J_{4422}^{23}$ & 3 &  3.609 &  0.2403 &  0.787 &  0.7875 &  0.8528 \\ 
45 & $J_{4422}^{66}$ & 3 &  3.6186 &  0.2422 &  0.7843 &  0.7848 &  0.8451 \\ 
46 & $J_{4422}^{81}$ & 3 &  3.5996 &  0.2397 &  0.7896 &  0.7904 &  0.8547 \\ 
47 & $J_{4422}^{71}$ & 3 &  3.5823 &  0.2471 &  0.7944 &  0.7945 &  0.8574 \\ 
48 & $J_{4422}^{88}$ & 2 &  2.616 &  0.2468 &  0.7851 &  0.7851 &  0.8454 \\ 
49 & $J_{4422}^{33}$ & 3 &  3.6151 &  0.2423 &  0.7853 &  0.7857 &  0.8516 \\ 
50 & $J_{4422}^{54}$ & 2 &  2.393 &  0.2307 &  0.8513 &  0.8539 &  0.8783 \\ 
51 & $J_{4422}^{111}$ & 2 &  2.7576 &  0.2455 &  0.7674 &  0.7676 &  0.856 \\ 
52 & $J_{4422}^{118}$ & 3 &  3.5283 &  0.2295 &  0.8255 &  0.8286 &  0.8829 \\ 
53 & $J_{4422}^{14}$ & 3 &  3.6023 &  0.2424 &  0.8059 &  0.8063 &  0.8682 \\ 
54 & $J_{4422}^{37}$ & 2 &  2.3909 &  0.2278 &  0.8648 &  0.8671 &  0.8942 \\ 
55 & $J_{4422}^{20}$ & 2 &  2.5356 &  0.2353 &  0.8236 &  0.8251 &  0.8624 \\ 
56 & $N_{4422}^{10}$ & 2 &  2.5 &  0.25 &  0.8333 &  0.8333 &  0.8633 \\ 
57 & $N_{4422}^4$ & 2 &  2.3956 &  0.2373 &  0.8634 &  0.8646 &  0.8922 \\ 
58 & $J_{4422}^9$ & 2 &  2.3423 &  0.222 &  0.8796 &  0.8835 &  0.8927 \\ 
59 & $J_{4422}^{82}$ & 2 &  2.7308 &  0.241 &  0.79 &  0.7908 &  0.8684 \\ 
60 & $J_{4422}^{67}$ & 2 &  2.6731 &  0.2455 &  0.8034 &  0.8035 &  0.8651 \\ 
61 & $J_{4422}^{51}$ & 3 &  3.6678 &  0.2419 &  0.8046 &  0.8051 &  0.8573 \\ 
62 & $N_{4422}^7$ & 3 &  3.6244 &  0.2425 &  0.815 &  0.8155 &  0.8645 \\ 
63 & $J_{4422}^2$ & 3 &  3.614 &  0.2353 &  0.8175 &  0.8189 &  0.8526 \\ 
64 & $J_{4422}^{102}$ & 3 &  3.6651 &  0.2385 &  0.8053 &  0.8062 &  0.8433 \\ 
65 & $J_{4422}^{87}$ & 3 &  3.6849 &  0.2419 &  0.8006 &  0.801 &  0.8545 \\ 
66 & $J_{4422}^{83}$ & 3 &  3.6692 &  0.2414 &  0.8043 &  0.8048 &  0.8571 \\ 
67 & $J_{4422}^{112}$ & 2 &  2.6248 &  0.2352 &  0.8149 &  0.8163 &  0.8607 \\ 
68 & $J_{4422}^{94}$ & 2 &  2.5686 &  0.238 &  0.8287 &  0.8298 &  0.8707 \\ 
69 & $J_{4422}^{24}$ & 3 &  3.6943 &  0.2432 &  0.7984 &  0.7987 &  0.8529 \\ 
70 & $J_{4422}^{35}$ & 3 &  3.6933 &  0.2464 &  0.7987 &  0.7987 &  0.8379 \\ 
71 & $J_{4422}^{36}$ & 3 &  3.6706 &  0.2393 &  0.8039 &  0.8048 &  0.8422 \\ 
72 & $J_{4422}^{22}$ & 5 &  5.8156 &  0.2426 &  0.7862 &  0.7866 &  0.8637 \\ 
73 & $J_{4422}^{61}$ & 5 &  5.8176 &  0.2495 &  0.7858 &  0.7858 &  0.8696 \\ 
74 & $J_{4422}^{27}$ & 3 &  3.9643 &  0.25 &  0.7568 &  0.7568 &  0.8514 \\ 
75 & $J_{4422}^{72}$ & 4 &  4.7878 &  0.2401 &  0.792 &  0.7929 &  0.8671 \\ 
76 & $J_{4422}^{41}$ & 4 &  4.7596 &  0.2356 &  0.798 &  0.7993 &  0.8606 \\ 
77 & $J_{4422}^{76}$ & 4 &  4.8291 &  0.2413 &  0.7835 &  0.784 &  0.8616 \\ 
78 & $J_{4422}^{62}$ & 4 &  4.75 &  0.25 &  0.8 &  0.8 &  0.861 \\ 
79 & $J_{4422}^{106}$ & 4 &  4.8382 &  0.2415 &  0.7816 &  0.7821 &  0.8604 \\ 
80 & $J_{4422}^{126}$ & 4 &  4.8024 &  0.2441 &  0.789 &  0.7894 &  0.8649 \\ 
81 & $J_{4422}^{77}$ & 4 &  4.8406 &  0.2408 &  0.7811 &  0.7817 &  0.8601 \\ 
82 & $J_{4422}^{116}$ & 2 &  2.7652 &  0.2429 &  0.7968 &  0.7972 &  0.8632 \\ 
83 & $J_{4422}^{50}$ & 3 &  3.8556 &  0.2438 &  0.7781 &  0.7784 &  0.8503 \\ 
84 & $J_{4422}^{16}$ & 4 &  4.5674 &  0.2305 &  0.8409 &  0.8434 &  0.8897 \\ 
85 & $J_{4422}^{19}$ & 4 &  4.6742 &  0.25 &  0.8165 &  0.8165 &  0.8719 \\ 
86 & $J_{4422}^4$ & 4 &  4.6862 &  0.2452 &  0.8138 &  0.814 &  0.8702 \\ 
87 & $J_{4422}^{42}$ & 2 &  2.6012 &  0.2362 &  0.8331 &  0.8342 &  0.8777 \\ 
88 & $J_{4422}^{90}$ & 4 &  4.8398 &  0.2449 &  0.7813 &  0.7815 &  0.8444 \\ 
89 & $J_{4422}^{58}$ & 4 &  4.8814 &  0.2457 &  0.7729 &  0.773 &  0.8386 \\ 
90 & $J_{4422}^{17}$ & 3 &  3.4288 &  0.2364 &  0.8749 &  0.8756 &  0.888 \\ 
91 & $J_{4422}^{34}$ & 2 &  2.4075 &  0.2377 &  0.8804 &  0.8808 &  0.9016 \\ 
92 & $J_{4422}^{121}$ & 3 &  3.5971 &  0.2274 &  0.834 &  0.8374 &  0.8713 \\ 
93 & $J_{4422}^{59}$ & 2 &  2.427 &  0.2372 &  0.8754 &  0.876 &  0.8848 \\ 
94 & $J_{4422}^{31}$ & 2 &  2.2133 &  0.2017 &  0.9336 &  0.9443 &  0.9462 \\ 
95 & $J_{4422}^7$ & 2 &  2.3642 &  0.2304 &  0.8917 &  0.8942 &  0.8859 \\ 
96 & $J_{4422}^8$ & 2 &  2.2657 &  0.1837 &  0.9186 &  0.9333 &  0.9254 \\ 
97 & $J_{4422}^{30}$ & 2 &  2.2459 &  0.1832 &  0.9242 &  0.9443 &  0.937 \\ 
98 & $J_{4422}^{11}$ & 2 &  2.2229 &  0.1865 &  0.9309 &  0.9428 &  0.9355 \\ 
99 & $J_{4422}^{110}$ & 5 &  5.9457 &  0.2447 &  0.7746 &  0.7748 &  0.8602 \\ 
100 & $J_{4422}^{70}$ & 5 &  5.9539 &  0.2411 &  0.7731 &  0.7737 &  0.8595 \\ 
101 & $J_{4422}^{93}$ & 5 &  5.9627 &  0.2423 &  0.7715 &  0.7719 &  0.8585 \\ 
102 & $J_{4422}^{107}$ & 5 &  5.934 &  0.2478 &  0.7768 &  0.7768 &  0.8584 \\ 
103 & $J_{4422}^{45}$ & 3 &  3.7572 &  0.2378 &  0.811 &  0.812 &  0.8682 \\ 
104 & $J_{4422}^{47}$ & 4 &  4.7645 &  0.2408 &  0.8096 &  0.8101 &  0.87 \\ 
105 & $J_{4422}^{28}$ & 4 &  4.75 &  0.25 &  0.8125 &  0.8125 &  0.861 \\ 
106 & $J_{4422}^{86}$ & 3 &  3.75 &  0.25 &  0.8125 &  0.8125 &  0.8571 \\ 
107 & $J_{4422}^{48}$ & 4 &  4.745 &  0.2409 &  0.8135 &  0.8142 &  0.8697 \\ 
108 & $J_{4422}^{15}$ & 3 &  3.6133 &  0.2433 &  0.8412 &  0.8417 &  0.8664 \\ 
109 & $J_{4422}^{122}$ & 2 &  2.6275 &  0.2381 &  0.8382 &  0.8391 &  0.8732 \\ 
110 & $N_{4422}^3$ & 3 &  3.6093 &  0.2425 &  0.8421 &  0.8427 &  0.8671 \\ 
111 & $N_{4422}^2$ & 3 &  3.6135 &  0.2338 &  0.8412 &  0.8429 &  0.8673 \\ 
112 & $J_{4422}^{129}$ & 3 &  4.0523 &  0.2434 &  0.7688 &  0.7692 &  0.8515 \\ 
113 & $J_{4422}^{80}$ & 4 &  4.9051 &  0.2464 &  0.7945 &  0.7946 &  0.8618 \\ 
114 & $J_{4422}^{64}$ & 4 &  4.9167 &  0.2466 &  0.7924 &  0.7926 &  0.8695 \\ 
115 & $J_{4422}^{68}$ & 4 &  5.0179 &  0.2416 &  0.7747 &  0.7753 &  0.8586 \\ 
116 & $J_{4422}^{78}$ & 3 &  3.8134 &  0.2406 &  0.8114 &  0.812 &  0.8705 \\ 
117 & $S_{242}^{51}$ & 5 &  6.0135 &  0.2417 &  0.7754 &  0.776 &  0.853 \\ 
118 & $J_{4422}^{99}$ & 3 &  3.8264 &  0.2388 &  0.809 &  0.8099 &  0.8587 \\ 
119 & $S_{242}^{52}$ & 4 &  4.8704 &  0.2364 &  0.8008 &  0.8021 &  0.8456 \\ 
120 & $J_{4422}^{124}$ & 4 &  4.941 &  0.2455 &  0.7881 &  0.7882 &  0.8474 \\ 
121 & $J_{4422}^{21}$ & 4 &  4.5441 &  0.2397 &  0.8655 &  0.8662 &  0.8924 \\ 
122 & $J_{4422}^{127}$ & 3 &  3.6147 &  0.2269 &  0.8506 &  0.8537 &  0.8683 \\ 
123 & $J_{4422}^5$ & 3 &  3.5007 &  0.2404 &  0.8749 &  0.8754 &  0.8734 \\ 
124 & $N_{4422}^6$ & 3 &  3.5971 &  0.2496 &  0.8543 &  0.8543 &  0.8546 \\ 
125 & $J_{4422}^{46}$ & 5 &  5.9717 &  0.2436 &  0.7942 &  0.7946 &  0.8661 \\ 
126 & $J_{4422}^{108}$ & 5 &  5.9676 &  0.2403 &  0.7949 &  0.7955 &  0.8581 \\ 
127 & $J_{4422}^{89}$ & 5 &  6.0036 &  0.239 &  0.7889 &  0.7898 &  0.8631 \\ 
128 & $J_{4422}^{96}$ & 3 &  3.9417 &  0.2413 &  0.7993 &  0.7999 &  0.8642 \\ 
129 & $J_{4422}^{117}$ & 5 &  5.9721 &  0.2447 &  0.7941 &  0.7944 &  0.866 \\ 
130 & $J_{4422}^{39}$ & 4 &  4.8265 &  0.2306 &  0.8194 &  0.8221 &  0.8793 \\ 
131 & $J_{4422}^{53}$ & 4 &  4.7994 &  0.2428 &  0.8243 &  0.8246 &  0.8653 \\ 
132 & $J_{4422}^{57}$ & 4 &  4.8053 &  0.2415 &  0.8232 &  0.8237 &  0.8646 \\ 
133 & $J_{4422}^{55}$ & 3 &  3.7066 &  0.2393 &  0.8415 &  0.8422 &  0.8752 \\ 
134 & $J_{4422}^{56}$ &  4 &  4.7491 &  0.2419 &  0.8335 &  0.834 &  0.872 \\ 
135 & $J_{4422}^3$ & 4 &  4.7249 &  0.2448 &  0.838 &  0.8382 &  0.8647 \\ 
136 & $J_{4422}^{49}$ & 4 &  4.8089 &  0.2392 &  0.8226 &  0.8234 &  0.8644 \\ 
137 & $N_{4422}^9$ & 4 &  4.7399 &  0.2445 &  0.8352 &  0.8354 &  0.8626 \\ 
138 & $J_{4422}^{98}$ & 5 &  6.1497 &  0.2396 &  0.7767 &  0.7776 &  0.8643 \\ 
139 & $J_{4422}^{85}$ & 4 &  4.9763 &  0.2362 &  0.8038 &  0.805 &  0.8694 \\ 
140 & $J_{4422}^{79}$ & 5 &  6.0156 &  0.2396 &  0.7975 &  0.7982 &  0.8529 \\ 
141 & $J_{4422}^{119}$ & 5 &  5.8489 &  0.2421 &  0.8249 &  0.8253 &  0.8714 \\ 
142 & $J_{4422}^{125}$ & 5 &  6 &  0.25 &  0.8 &  0.8 &  0.8541 \\ 
143 & $J_{4422}^{105}$ & 5 &  6.0742 &  0.2418 &  0.7883 &  0.7888 &  0.8465 \\ 
144 & $J_{4422}^{65}$ & 4 &  5.111 &  0.2466 &  0.7826 &  0.7827 &  0.8484 \\ 
145 & $J_{4422}^{113}$ & 3 &  3.8195 &  0.2399 &  0.83 &  0.8306 &  0.8698 \\ 
146 & $J_{4422}^{101}$ & 5 &  6.0296 &  0.2429 &  0.7953 &  0.7956 &  0.8439 \\ 
147 & $J_{4422}^{128}$ & 5 &  6.0096 &  0.2497 &  0.7985 &  0.7985 &  0.8531 \\ 
148 & $N_{4422}^{11}$ & 3 &  3.4917 &  0.2246 &  0.8905 &  0.8947 &  0.8914 \\ 
149 & $J_{4422}^{13}$ & 3 &  3.5629 &  0.2422 &  0.8766 &  0.8772 &  0.8615 \\ 
150 & $J_{4422}^{91}$ & 7 &  8.2993 &  0.2422 &  0.7659 &  0.7663 &  0.858 \\ 
151 & $J_{4422}^{92}$ & 4 &  5.0648 &  0.2405 &  0.7997 &  0.8002 &  0.8622 \\ 
152 & $J_{4422}^{52}$ & 3 &  4.0999 &  0.2442 &  0.7944 &  0.7947 &  0.8586 \\ 
153 & $J_{4422}^{115}$ & 3 &  3.9802 &  0.2394 &  0.8126 &  0.8134 &  0.8687 \\ 
154 & $J_{4422}^{38}$ & 4 &  4.9295 &  0.2342 &  0.8205 &  0.8225 &  0.8693 \\ 
155 & $J_{4422}^{40}$ & 6 &  6.902 &  0.2441 &  0.833 &  0.8333 &  0.883 \\ 
156 & $J_{4422}^{84}$ & 4 &  5.067 &  0.2407 &  0.8083 &  0.8089 &  0.8697 \\ 
157 & $J_{4422}^{95}$ & 5 &  5.9418 &  0.2401 &  0.8269 &  0.8277 &  0.8697 \\ 
158 & $J_{4422}^{120}$ & 3 &  3.7931 &  0.2385 &  0.8502 &  0.8509 &  0.8899 \\ 
159 & $J_{4422}^{18}$ & 5 &  5.7018 &  0.2416 &  0.8651 &  0.8653 &  0.8892 \\ 
160 & $J_{4422}^{29}$ & 5 &  6.0246 &  0.2461 &  0.8145 &  0.8146 &  0.8515 \\ 
161 & $J_{4422}^{114}$ & 5 &  6.0653 &  0.2463 &  0.8086 &  0.8087 &  0.8562 \\ 
162 & $N_{4422}^8$ & 5 &  6.0189 &  0.2408 &  0.8154 &  0.816 &  0.8614 \\ 
163 & $J_{4422}^{103}$ & 5 &  6.052 &  0.2385 &  0.8105 &  0.8114 &  0.8491 \\ 
164 & $N_{4422}^1$ & 5 &  6.0076 &  0.2467 &  0.8171 &  0.8171 &  0.8622 \\ 
165 & $J_{4422}^{123}$ & 5 &  6.0641 &  0.2422 &  0.8088 &  0.8092 &  0.8475 \\ 
166 & $J_{4422}^{10}$ & 3 &  3.3738 &  0.2138 &  0.9233 &  0.9281 &  0.9155 \\ 
167 & $J_{4422}^{12}$ & 3 &  3.4198 &  0.25 &  0.9147 &  0.9147 &  0.8893 \\ 
168 & $J_{4422}^{73}$ & 4 &  5.1205 &  0.2478 &  0.8091 &  0.8092 &  0.8714 \\ 
169 & $J_{4422}^{100}$ & 6 &  7.2675 &  0.2416 &  0.7894 &  0.7899 &  0.856 \\ 
170 & $N_{4422}^{12}$ & 7 &  8.325 &  0.2431 &  0.7905 &  0.7909 &  0.8665 \\ 
171 & $J_{4422}^{97}$ & 4 &  5.1584 &  0.2477 &  0.8119 &  0.8119 &  0.8657 \\ 
172 & $N_{4422}^5$ & 7 &  8.4377 &  0.2403 &  0.785 &  0.7857 &  0.8516 \\ 
173 & $J_{4422}^{104}$ & 6 &  7.5876 &  0.2398 &  0.7836 &  0.7844 &  0.8574 \\ 
174 & $J_{4422}^{109}$ & 9 &  10.7261 &  0.2417 &  0.7766 &  0.7771 &  0.8552 \\ 
\hline
\end{longtabu}

\end{document}